\documentclass[aps,prb,twocolumn,groupedaddress,dvipdfmx]{revtex4-1}
\usepackage{graphicx}
\usepackage{dcolumn}
\usepackage{bm}
\usepackage{amsmath}
\usepackage{txfonts}
\usepackage{url}
\usepackage{longtable}

\begin{document}

\title[calcHamakerConst]
  {Method for the calculation of the Hamaker constants of organic materials by the Lifshitz macroscopic approach with DFT}

\author{Hideyuki Takagishi$^{1,*}$}
\author{Takashi Masuda$^{1}$}
\author{Tatsuya Shimoda$^{1}$}
\author{Ryo Maezono$^{2,3}$}
\author{Kenta Hongo$^{3,4,5}$}

\affiliation{$^{1}$
  School of Materials Science, JAIST, Asahidai 1-1, Nomi, Ishikawa,
  923-1292, Japan
}

\affiliation{$^{2}$
  School of Information Science, JAIST, Asahidai 1-1, Nomi, Ishikawa,
  923-1292, Japan
}

\affiliation{$^{3}$
  Computational Engineering Applications Unit, RIKEN, 
  2-1 Hirosawa, Wako, Saitama 351-0198, Japan
}

\affiliation{$^{3}$
  Research Center for Advanced Computing Infrastructure, JAIST,
  Asahidai 1-1, Nomi, Ishikawa 923-1292, Japan
}
\affiliation{$^{4}$
  Center for Materials Research by Information Integration,
  Research and Services Division of Materials Data and
  Integrated System, National Institute for Materials Science,
  Tsukuba 305-0047, Japan
}
\affiliation{$^{5}$
  PRESTO, Japan Science and Technology Agency, 4-1-8 Honcho,
  Kawaguchi-shi, Saitama 322-0012, Japan
}

\affiliation{$^{*}$
    h-takagishi@jaist.ac.jp
}

\begin{abstract}
The Hamaker constants, which are coefficients providing quantitative information on intermolecular forces, were calculated for a number of different materials according to the Lifshitz theory via simple DFT calculations without any experimental measurements being performed. 
The physical properties (polarizability, dipole moment, molecular volume, and vibrational frequency) of organic molecules were calculated using the B3LYP density functional and the aug-cc-pVDZ basis set. 
Values for the Hamaker constants were obtained using the approximation of the Lorentz-Lorenz equation and Onsager's equation with these properties. 
It was found that, in the case of `non-associative' materials, like hydrocarbons, ethers, ketones, aldehydes, carboxylic acids, esters, nitriles, and hydrosilanes, and halides, the calculated Hamaker constants were similar in value to their experimentally determined counterparts. Moreover, with this calculation method, it is easy to create the molecular model and the CPU time can be shortened.
\end{abstract}

\maketitle
\newpage

\section{Introduction}
The Hamaker constant is a coefficient accounting for the van der Waals interaction between two materials, and it has a strong correlation with various physical phenomena, such as liquid wettability, adhesion, friction, adsorption, colloidal stability, polymer flow, and deformation. \cite{1937HAM, 2011ISR, 2006MAS}
Various methods for the calculation of the Hamaker constant have been developed by many researchers: microscopic approximation,\cite{1930EIS, 1931SLA, 1932KIR, 1937NEU, 1969TAB}
macroscopic approximation,\cite{1956LIF, 1980HOU} 
calculation from the dielectric constant,\cite{1969TAB, 1970GRE}
calculation from the surface energy,\cite{1973IS2, 1969OWE}
calculation from the critical coagulation concentration of colloids, 
calculation from the materials' rheological characteristics,\cite{1963SHE, 1968HUN, 1971FRI}
and direct measurement by scanning probe microscopy.\cite{1999CAP, 2005BUT}
In recent years, values for the Hamaker constants and the intermolecular forces have been reported to be obtainable by way of quantum chemical calculations, in the absence of experimental measurements.\cite{2017HON, 2010ZHA, 2007RAJ, 2002BHA}
These methods correspond to the microscopic approximation. 
The simplest calculation method in the microscopic approximation is the London/Hamaker approach, which is based on the London formula.\cite{1930EIS}
However, the London/Hamaker approach can only be used to perform calculations on small, non-polar molecules.\cite{2011ISR}
Furthermore, although more advanced calculation methods have been devised, these approaches are more complicated than London/Hamaker's, so they are difficult to implement. 
A careful selection of the theory and the molecular geometry are necessary to make appropriate calculations for the estimation of intermolecular forces. 
As is well known, the popular density functional B3LYP cannot be utilized to calculate intermolecular forces accurately; by contrast, the density functional including the dispersion force correction term (e.g., wB97XD or M06-2X) or a method other than DFT (e.g., MP2 or CCSD(T)) are required to perform the mentioned calculations.\cite{2008CHA, 2008ZHA,2004GRI}
In addition, it is necessary to create molecular models of various aggregate geometries.\cite{2017HON}
The implementation of these methods is also quite demanding in terms of CPU time.

Therefore, in this study, we aimed to develop an easily applicable and rapid method to calculate the values of the Hamaker constants of various organic materials via the Lifshitz macroscopic approach, a method that is about as demanding as the London/Hamaker approach when it comes to CPU time.
According to the Lifshitz macroscopic theory,\cite{1956LIF} the Hamaker constant can be calculated from the frequency dependence of the dielectric function. 
Conventionally, the refractive index, the relative dielectric constant, and the infrared spectrum of the organic materials are experimentally measured and used to calculate the Hamaker constant. 
In the present study, instead of conducting experimental measurements, the refractive index, relative dielectric constant, and infrared spectrum of a given material were estimated by performing simple DFT calculations. 
In particular, the value of the refractive index was approximated employing the Lorentz-Lorenz equation\cite{1880LOR, 1880LO2} and utilizing the calculated polarizability and the calculated molecular volume of the single molecule. 
The relative dielectric constant was approximated using Onsager's equation\cite{1936ONS} and employing the calculated dipole moment, the calculated polarizability, and the calculated molecular volume of the single molecule. 
Finally, the infrared spectrum was approximated utilizing the vibrational frequencies of the single molecule. 
Evidence indicated that the Hamaker constants calculated by this method were similar to those calculated based on experimental results, except in the case of ``associative'' materials (\textit{vide infra}) like alcohols. 
Since this method only required a single molecular model for organic materials, the values of the Hamaker constants could be calculated with a general density functional like B3LYP without any correction. 
Notably, the creation of a molecular model was easy to achieve, and the CPU time necessary to perform the calculations was short.

\section{Theory}
In Lifshitz theory, the Hamaker constant $A_{132}$ between material 1 and material 2 interacting over material 3 can be calculated from the frequency dependence of the dielectric function $\epsilon$ using Eq(\ref{eq.Lifshitz}):

\begin{equation}
\begin{split}
A_{132} & =  \frac{3}{4}k_BT \left(\frac{\epsilon_1-\epsilon_3}{\epsilon_1+\epsilon_3}\right)\left(\frac{\epsilon_2-\epsilon_3}{\epsilon_2+\epsilon_3}\right) \\ 
& +\frac{3h}{4\pi} \int_{\nu_1}^{\infty}\left(\frac{\epsilon_1(i\nu)-\epsilon_3(i\nu)}{\epsilon_1(i\nu)+\epsilon_3(i\nu)}\right)\left(\frac{\epsilon_2(i\nu)-\epsilon_3(i\nu)}{\epsilon_2(i\nu)+\epsilon_3(i\nu)}\right)d \nu ,
\end{split}
\label{eq.Lifshitz}
\end{equation}
where $k_B$ is the Boltzmann constant, $T$ is the absolute temperature, and $h$ is the Planck constant.
Eq(\ref{eq.Lifshitz_appr}) is an approximated formula for Eq(\ref{eq.Lifshitz}) that is often used for organic\cite{2009MAS} and inorganic\cite{1996BER} materials.

\begin{equation}
A_{132}=\frac{3k_BT}{2} \sum_{n=0}^{\infty}\,^{\prime} \sum_{s=1}^{\infty}\frac{(\Delta _{13} \Delta_{23})^s}{s^3} .
\label{eq.Lifshitz_appr}
\end{equation}
The prime symbol that refers to the first summation indicates that the value is multiplied by 0.5 when $n$ = 0 (the static contribution). 
Each parameter in Eq(\ref{eq.Lifshitz_appr}) is calculated employing Eq (\ref{eq.deltakj})-Eq (\ref{eq.ixi}).

\begin{equation}
\Delta_{kj}=\frac{\epsilon_k(i\xi_n)-\epsilon_j(i\xi_n)}{\epsilon_k(i\xi_n)+\epsilon_j(i\xi_n)}
\label{eq.deltakj}
\end{equation}

\begin{equation}
\xi_n=n\frac{4\pi ^2k_BT}{h}
\end{equation}

\begin{equation}
\epsilon(i\xi_n)=1+\frac{C_{UV}}{1+(\xi_n/\omega_{UV})^2}+\frac{C_{IR}}{1+(\xi_n/\omega_{IR})^2} .
\label{eq.ixi}
\end{equation}

By this approach, the Hamaker constant $A_{132}$ can be calculated from the optical parameters $C_{UV}$, $\omega_{UV}$, $C_{IR}$, and $\epsilon_{IR}$ of each material. 
In particular, parameters $C_{UV}$ and $\epsilon_{UV}$ can be obtained from the frequency dependence of the refractive index via the Cauchy equation Eq(\ref{eq.Cauchy}):\cite{1980HOU}

\begin{equation}
n^2-1=\frac{(n^2-1)\omega^2}{\omega_{UV}^2}+C_{UV}
\label{eq.Cauchy}
\end{equation}

\begin{equation}
\omega=\frac{2\pi c}{\lambda} ,
\label{eq.convOmega}
\end{equation}
where $c$ is speed of light in the vacuum and $\lambda$ is the wavelength of the incident light. 
The mathematical expression of Eq(\ref{eq.Cauchy}) implies that the slope and the intercept of the linear plot of $n^2-1$ versus $(n^2 - 1)\omega^2$ correspond to $1/\omega_{UV}^2$ and $C_{UV}$, respectively.
$C_{IR}$ can be calculated from the relative dielectric constant ($\epsilon_r$) and $C_{UV}$ employing Eq(\ref{eq.CIR}):

\begin{equation}
C_{IR} + C_{UV} + 1 = \epsilon_r ,
\label{eq.CIR}
\end{equation}
$\omega_{IR}$ can be obtained by collecting the wavenumber ($\nu$) of the strongest absorption peak in the infrared spectrum and plugging its value into Eq(\ref{eq.OIR}):

\begin{equation}
\omega_{IR}=2 \pi c \nu .
\label{eq.OIR}
\end{equation}

In this study, the values for the Hamaker constants for identical organic materials in air ($A_{11}$) were calculated for comparison with the values for the same parameter obtained experimentally. 
Notably, $A_{11}$ can be calculated utilizing Eq(\ref{eq.Lifshitz_appr})-Eq(\ref{eq.ixi}) keeping into account that material 1 = material 2 and material 3 = air.
The value of the refractive index $n$ can be approximated using the Lorentz-Lorenz equation (Eq(\ref{eq.Lorentz-Lorenz})):\cite{1880LOR, 1880LO2}

\begin{equation}
\frac{n^2-1}{n^2+2}=\frac{2\pi \alpha}{3V} ,
\label{eq.Lorentz-Lorenz}
\end{equation}
where $\alpha$ is the material's polarizability and $V$ is its molecular volume. 
In the case of a crystalline material, it is necessary to consider the regularity, whereas in the case of an amorphous material it is unnecessary to consider. 
Since the organic materials taken into consideration in this study were liquid, and thus amorphous in nature, Eq(\ref{eq.Lorentz-Lorenz}) was used without correction.
The value of the relative dielectric constant ($\epsilon_r$) can be approximated using Onsager's equation (Eq(\ref{eq.Onsager})):\cite{1936ONS}

\begin{equation}
\frac{(\epsilon_r-n^2)(2\epsilon_r+n^2)M_w}{\epsilon_r(n^2+2)^2d}=\frac{N_A \mu^2}{9 \epsilon_0 k_B T} ,
\label{eq.Onsager}
\end{equation}
where $\epsilon_r$ is the dielectric constant of vacuum, $M_w$ is the material's molecular weight, $d$ is its density, $\mu$ is its dipole moment, and $N_A$ is the Avogadro constant. 
Notably, the value of $d$ can be calculated from that of $V$ using Eq(\ref{eq.dFromV}):

\begin{equation}
d= \frac{M_w}{N_A  V} .
\label{eq.dFromV}
\end{equation}
.

\section{Methods and Calculations}
All the quantum chemical calculations were performed using the Gaussian 16 software package\cite{G16} with Becke's three-parameter nonlocal exchange functional along with the Lee-Yang-Parr nonlocal correlation functional (B3LYP),\cite{1993BEC, 1988LEE} unless otherwise stated. 
In this study, polarizability ($\alpha$), molecular volume ($V$), refractive index ($n$), dipole moment ($\mu$), relative dielectric constant ($\epsilon_r$), and Hamaker constant $A_{11}$ were calculated for common organic materials that are in liquid phase at room temperature. 
Specifically, the organic materials for which the mentioned parameters were calculated in this study were hydrocarbons, ethers, ketones, aldehydes, carboxylic acids, esters, nitriles, hydrosilanes, halides, alcohols, amines, and amides. The materials having plural types of functional groups were excluded from the calculation. 
All the physical properties of single molecules were calculated after the relevant molecular structure had been optimized. 
The mean absolute relative error (MARE) of a calculated parameter was calculated as follows to evaluate:
\[
MARE=\frac{1}{N} \sum \frac{|x_{calcd}-x_{expl}|}{x_{expl}} ,
\]
where $x_{expl}$ is the experimental value of the parameter being evaluated, $x_{calcd}$ is its calculated value, and $N$ is the number of samples. 
Please note that the experimental values for the parameters were taken from the literature.\cite{2007CRC, webbook, pubchem}

\subsection{Selection of basis set}
The values of the polarizability of common materials were calculated, and the relevant MARE values were compared to enable us to select the best basis set. 
The following basis sets were used with the B3LYP method: 6-31++G(d,p), 6-311++G(d,p), aug-cc-pVDZ, and aug-cc-pVTZ. 
The small molecules for which the relevant parameters were calculated to act as a benchmark were 
$\mathrm{H_2}$, $\mathrm{N_2}$, $\mathrm{CH_4}$, $\mathrm{C_2H_2}$, $\mathrm{C_2H_4}$, $\mathrm{C_2H_6}$, $\mathrm{C_3H_6}$, $\mathrm{C_6H_6}$, $\mathrm{HF}$, $\mathrm{HCl}$, $\mathrm{HBr}$, $\mathrm{Cl_2}$, $\mathrm{CCl_4}$, $\mathrm{NH_3}$, $\mathrm{CH_3NH_2}$, $\mathrm{CO}$, $\mathrm{CO_2}$, $\mathrm{COS}$, $\mathrm{HCHO}$, $\mathrm{CH_3CHO}$, $\mathrm{H_2O}$, $\mathrm{CH_3OH}$, $\mathrm{CH_3OCH_3}$, $\mathrm{H_2S}$, $\mathrm{CS_2}$, $\mathrm{SO_2}$, $\mathrm{SiH_4}$, $\mathrm{SiF_4}$ and $\mathrm{Si_2H_6}$.
In order to evaluate, second-order M{\o}ller-Plesset perturbation theory (MP2),\cite{1934MOL} which is expected to afford parameter values characterized by relatively good accuracy, was also used.

\subsection{Calculation of the polarizability and the dipole moment}
The values for the molecules' polarizability ($\alpha$) and dipole moment ($\mu$) were calculated using the Gaussian keyword "polar" with B3LYP/aug-cc-pVDZ, which had been selected based on the procedure described above. 
The calculation results were compared with the relevant experimental values, and these results were used for the following calculation.

\subsection{Calculation and correction of the density value}
The molecular volume of a given single molecule was calculated ten times using the Gaussian keyword "volume (tight)", and the average value obtained after these ten calculations was defined as $V_{calcd}$. 
This value for $V_{calcd}$ was then used to calculate the molecule's density ($d_{calcd}$) via Eq(\ref{eq.dFromV}). 
Notably, organic materials were classified into three groups: non-associative liquids (i.e. hydrocarbons, ethers, ketones, aldehydes, carboxylic acids, esters, nitriles, and hydrosilanes), associative liquids (i.e. alcohols, amines, and amides), and halides. 
The regression line $d_{expl} = a \cdot d_{calcd} + b$ was calculated employing the least squares method in each of the three classification groups. 
The values for the corrected density $d_{corr} = (d_{calcd}-b)/a$ were calculated for each material. 
The values for the corrected molecular volume ($V_{corr}$) were calculated from the relevant $d_{corr}$ values via Eq(\ref{eq.dFromV}), and the $V_{corr}$ values thus obtained were used for the following calculation.

\subsection{Calculation of the Hamaker constant}
First, the molecules' dynamic polarizability ($\alpha$) was calculated using the coupled perturbed Hartree-Fock calculation implemented in Gaussian16 at wavelengths ($\lambda$) equal to 450, 500, 550, 589, 600, 650, 700, 750, 800, 900, and 1000 nm, using the Gaussian keyword "polar cphf". 
The refractive index ($n$) was approximated via Eq(\ref{eq.Lorentz-Lorenz}) using the calculated values for $\alpha$ and $V_{corr}$.
Next, the values for $C_{UV}$ and $\omega_{UV}$ were obtained by conducting a Cauchy plot analysis. 
In particular, the following expression was plotted: (x, y) = ($(n^2 - 1)\omega^2$, $n^2 - 1$). 
The slope of the regression line of this plot corresponds to the value of $1/\omega_{UV}^2$, whereas the plot's intercept corresponds to the value of $C_{UV}$.
The values for the relative dielectric constant ($\epsilon_r$) were approximated using Eq(\ref{eq.Onsager}), based on the values for $\alpha$ and $V_{corr}$ calculated as described above.
The absolute temperature ($T$) was set to 293.15 K, the same value utilized in the experimental conditions. 
The values for $C_{IR}$ were calculated employing Eq(\ref{eq.CIR}) and using the calculated values for $\epsilon_r$ and $C_{UV}$.
The vibrational frequencies of the molecules were calculated using the Gaussian keyword "freq", which afforded the value for the wavenumber $\nu_{calcd}$ of the strongest peak.
The values for the calculated wavenumbers were then corrected to obtain $\nu_{corr}$ values, obtained via the following expression: by $\nu_{corr} = f \cdot \nu_{calcd}$, where $f$ is the scaling factor. 
In this study, a value for $f$ of 0.970 was utilized, as indicated for B3LYP/aug-cc-pVDZ by the National Institute of Standards and Technology.\cite{NIST} 
The corrected wavenumber ($\nu_{corr}$) was used to obtain the value of $\omega_{IR}$ by Eq(\ref{eq.OIR}).
The values for the Hamaker constant $A_{11}$ were calculated by Eq(\ref{eq.Lifshitz_appr}) using the values for $C_{UV}$, $\omega_{UV}$, $C_{IR}$, and $\omega_{IR}$, and utilizing the following integration ranges: $s$ = 1-9 and $n$ = 0-9999.

For comparison, the values for the Hamaker constant $A_{11}$ were also calculated by the London/Hamaker approach using Eq(\ref{eq.microscopic}):\cite{1937HAM, 1930EIS}

\begin{equation}
A_{11}=\pi^2 C_6 \rho^2 ,
\label{eq.microscopic}
\end{equation}
where $\rho$ is the number density (molecules per unit volume) and $C_6=3/4\alpha ^2 I$. 
Notably, $I$ , the vertical ionization potential, was calculated as the difference between the total electronic energy of the cationic and neutral molecule.

\section{Results and Discussion}
\subsection{Selection of the basis set}
The first step in this study was to select the most suitable basis set, because the basis set has a big influence on the accuracy of the calculation of a molecule's polarizability. 
In Table \ref{tab.MARE_common} are reported data that allowed us to determine how a particular basis set influenced the accuracy of the values obtained for a molecule’s polarizability, a parameter that reflects the change in the charge distribution within molecules as a consequence of external electric fields. 
Notably, as polarizability is strongly influenced by the behavior of electrons far from the nucleus, large basis sets that can calculate the spread of electrons are preferable to small basis sets.\cite{1998TSU, 1994CHA}
The values for polarizability calculated using the aug-cc-pVDZ or aug-cc-pVTZ basis sets are close to the corresponding experimental values. 
Although use of aug-cc-pVTZ required a longer CPU time than use of aug-cc-pVDZ, the differences in calculation accuracy were small between the two basis sets. 
The polarizability value accuracy as calculated by B3LYP was almost the same as that calculated by MP2, although the CPU time was very short. 
Based on these results, the B3LYP/aug-cc-pVDZ combination was adopted to perform the subsequent calculations.

\begin{table} [htbp]
\caption{Mean absolute relative error for the calculated polarizability values of 29 common organics}
\label{tab.MARE_common}
\begin{center}
\begin{tabular} { l c c } \hline 
 & B3LYP & MP2 \\ \hline
6-31++G(d,p) & 0.20  & 0.23  \\ 
6-311++G(d,p) & 0.20  & 0.22  \\ 
aug-cc-pVDZ & 0.06  & 0.07  \\ 
aug-cc-pVTZ & 0.05  & 0.07  \\ 
\hline
\end{tabular}
\end{center}
\end{table}

\subsection{Calculation of the polarizability and dipole moment values}
\begin{figure}[h]
\centering
\includegraphics[width=7cm]{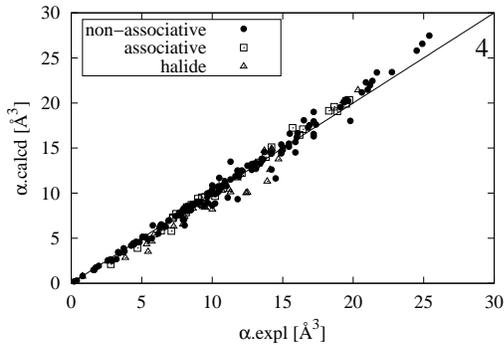}
\vspace{10mm}
\caption{Plot of the calculated values of polarizability ($\alpha_{calcd}$) versus their experimentally determined counterparts ($\alpha_{expl}$) for non-associative materials (131 compounds), associative materials (43 compounds), and halides (35 compounds).}
\label{fig.polar}
\end{figure}

\begin{figure}[h]
\centering
\includegraphics[width=7cm]{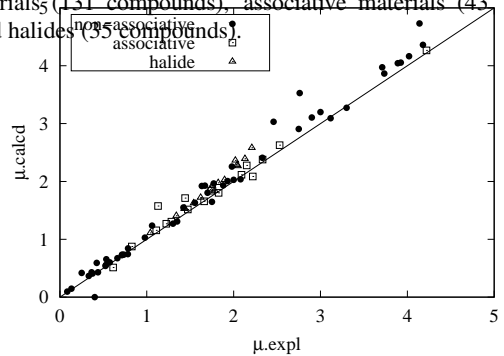}
\vspace{10mm}
\caption{Plot of the calculated values of the dipole moment ($\mu_{calcd}$) versus their experimentally determined counterparts ($\mu_{expl}$) for non-associative materials (42 compounds), associative materials (14 compounds), and halides (16 compounds).}
\label{fig.dipole}
\end{figure}

The values for the polarizability ($\alpha$) and dipole moment ($\mu$) were calculated for a total of 209 materials using the B3LYP/aug-cc-pVDZ combination. 
From the data reported in Figure \ref{fig.polar} can be evinced the correlation between the calculated values of the polarizability ($\alpha_{calcd}$) and the experimentally determined ones ($\alpha_{expl}$). 
In Figure \ref{fig.dipole} are reported data that reflect the correlation between the calculated values of the dipole moment ($\mu_{calcd}$) and the experimentally determined ones ($\mu_{expl}$). 
It was found that the parameter calculation error for most materials was less than 10\%, and it was possible to calculate the values of each of the two parameters with high precision for any type of material (Table \ref{tab.MARE_all}).

\subsection{Calculation of corrected density values}
\begin{figure}
\centering
\includegraphics[width=7cm]{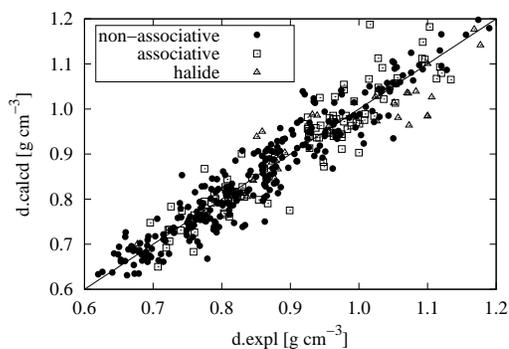}
\vspace{10mm}
\caption{Plot of the calculated values of the density ($d_{calcd}$) versus their experimentally determined counterparts ($d_{expl}$) for non-associative materials (278 compounds), associative materials (101 compounds), and halides (45 compounds).}
\label{fig.density}
\end{figure}

Several methods have been devised for calculating the molecular volume of a chemical, among them integrating the atom volume based on van der Waals radii\cite{1964BON, 1970SLO} and the use of a Monte Carlo-based integration. 
In this study, a Monte Carlo-based integration was implemented in Gaussian16 for this purpose. 
In this method, the area of space where the electron density is 0.001 $\mathrm{electrons/Bohr^3}$ is defined as the surface of the molecule. 
It has already been reported that the density of solid organic materials can be calculated with high accuracy by the described approach using B3LYP.\cite{2007RIC, 2007QIU, 2010BUL} 
In the case of liquid organic materials, in this study the density values calculated using B3LYP/aug-cc-pVDZ displayed a tendency to be uniformly larger than their experimental counterparts. 
Since the position of the molecules is not restricted in the actual liquid, the average spacing between molecules is presumed to be wider than it is in the calculation. 
Palomer \textit{et al.} use the regression line to correct the calculated density of ionic liquids.\cite{2007PAL} 
The same corrections were implemented in the calculation performed in this study. In particular, the regression line was obtained comparing the calculated values for the density of the various material groups (see below) with the experimental ones. 
Based on conventional experience, materials were classified into three groups: "non-associative materials" (hydrocarbons, ethers, ketones, aldehydes, esters, and hydrosilanes), "associative materials" (alcohols, amines and amides), or "halides". 
The expressions for the regression lines for each of the three groups were as follows: 
$d_{calcd} = 0.858 d_{expl} + 0.345$, $d_{calcd} = 0.728 d_{expl} + 0.428$, and $d_{calcd} = 1.062 d_{expl} + 0.199$, respectively. 
As can be evinced from the data reported in Figure \ref{fig.density} and Table \ref{tab.MARE_all}, the calculation error for most materials was under the 5\% mark.

\subsection{Calculation of Hamaker constant values via the Lifshitz and London/Hamaker approaches}
\begin{figure}
\centering
\includegraphics[width=7cm]{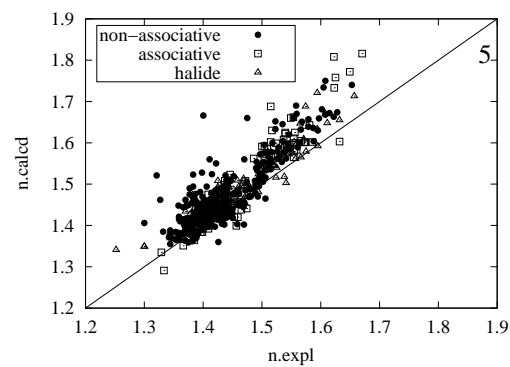}
\vspace{10mm}
\caption{Plot of the calculated values of the refractive index ($n_{calcd}$) versus their experimentally determined counterparts ($n_{expl}$) at $\lambda$ = 598 nm for non-associative materials (319 compounds), associative materials (128 compounds), and halides (60 compounds).}
\label{fig.refractiveIdx}
\end{figure}

\begin{figure}
\centering
\includegraphics[width=7cm]{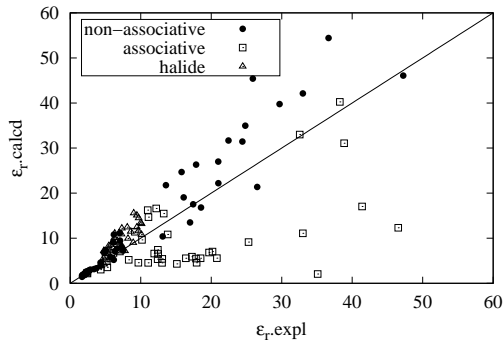}
\vspace{10mm}
\caption{Plot of the calculated values of the relative dielectric constant ($\epsilon_{r.calcd}$) versus their experimentally determined counterparts ($\epsilon_{r.expl}$) for non-associative materials (62 compounds), associative materials (47 compounds), and halides (37 compounds).}
\label{fig.epsilon}
\end{figure}

In Figure \ref{fig.refractiveIdx} are reported data reflecting the correlation between the calculated values for the refractive index ($n_{calcd}$) and those of the relevant experimentally determined counterparts ($n_{expl}$) at a wavelength of $\lambda$ = 598 nm. 
It has already been reported that by employing the B3LYP density functional the refractive index of organic materials can be accurately estimated.\cite{2006AND} 
In this study, the calculated values for the refractive index tended to be slightly lower than their experimentally determined counterparts, but the calculation error remained below the 5\% mark (Table \ref{tab.MARE_all}).
Approximate expressions for the calculation of the relative dielectric constant ($\epsilon_r$) have been proposed by Clausius-Mossotti,\cite{1850MOS, 1879CLA, 1939WIL} Debye,\cite{1929DEB} Onsager\cite{1936ONS}, and Kirkwood.\cite{1939KIR} 
In this study, Onsager's equation was utilized, which can be expected to produce results characterized by relatively good accuracy via a simple calculation. 
In Figure \ref{fig.epsilon} are reported data reflecting the correlation between the calculated values of the relative dielectric constant ($\epsilon_{r.calcd}$) and the corresponding experimental values ($\epsilon_{r.expl}$). 
The values calculated for non-associative materials and halides tended to be consistently slightly higher than their experimentally determined counterparts. 
This overestimation of the dielectric constant is influenced by the overestimation of the refractive index, as can be evinced from the data reported in Figure \ref{fig.refractiveIdx}. 
For some materials, the error in the calculated value of the dielectric constant is over 50\%, so in this case result accuracy is not so good. 
In the case of associative materials, the calculated values of the dielectric constant tended to be significantly lower than their experimentally determined counterparts. 
Although use of Kirkwood's equation has been reported to solve this problem,\cite{2007VAS} in order to use this equation, it is necessary to obtain a positional relationship with the proximity molecules. 
Given that the purpose of this study was to calculate the Hamaker constants by a simple calculation approach and that Kirkwood's approximation requires instead complicated calculations, Onsager's equation was utilized in this study.

\begin{figure}
\centering
\includegraphics[width=7cm]{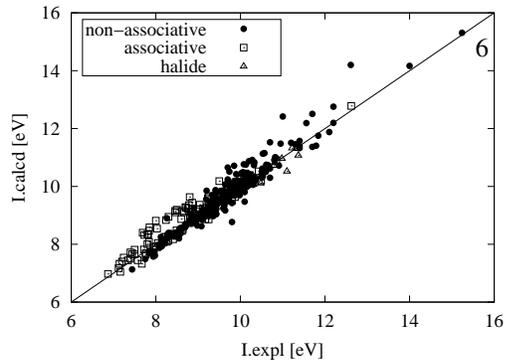}
\vspace{10mm}
\caption{Plot of the calculated values of the ionization potential ($I_{calcd}$) versus their experimentally determined counterparts ($I_{expl}$) for non-associative materials (278 compounds), associative materials (106 compounds) and halides (21 compounds).}
\label{fig.ionizationPotential}
\end{figure}

\begin{figure}[h]
\centering
\includegraphics[width=7cm]{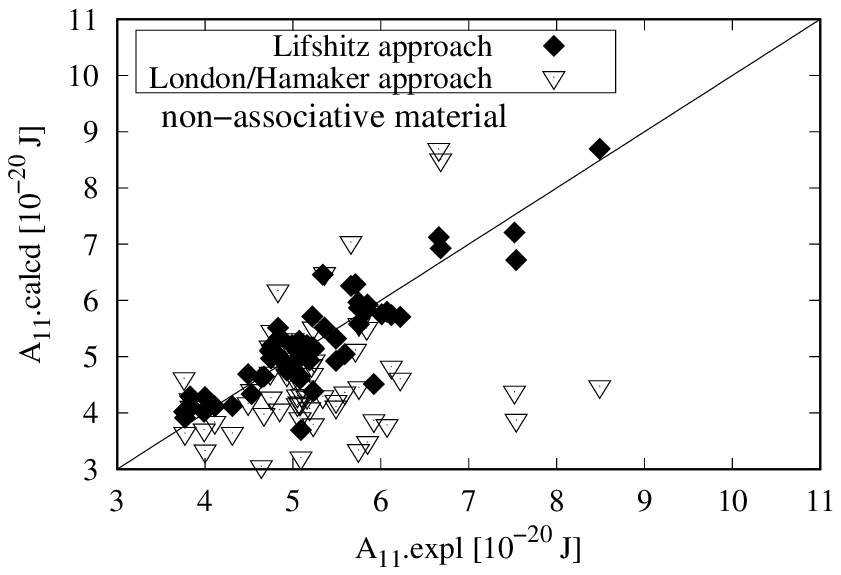}

\includegraphics[width=7cm]{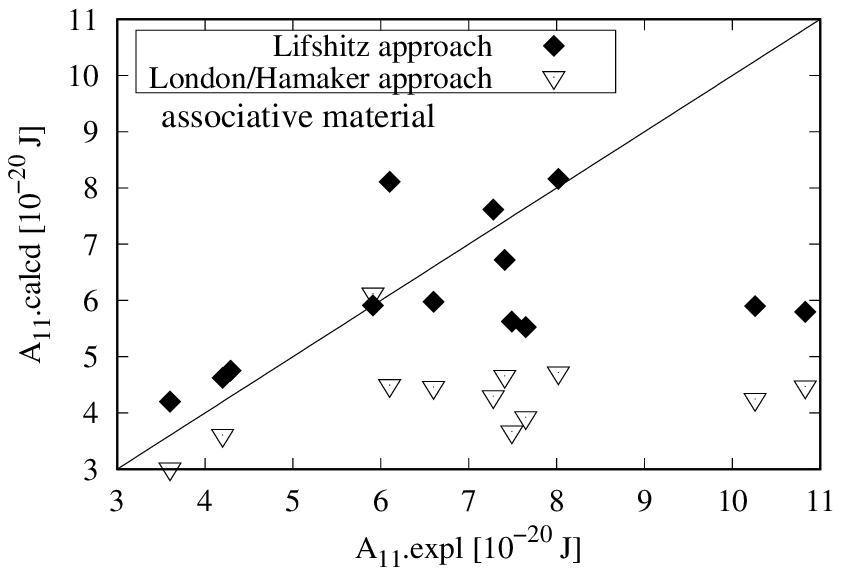}

\includegraphics[width=7cm]{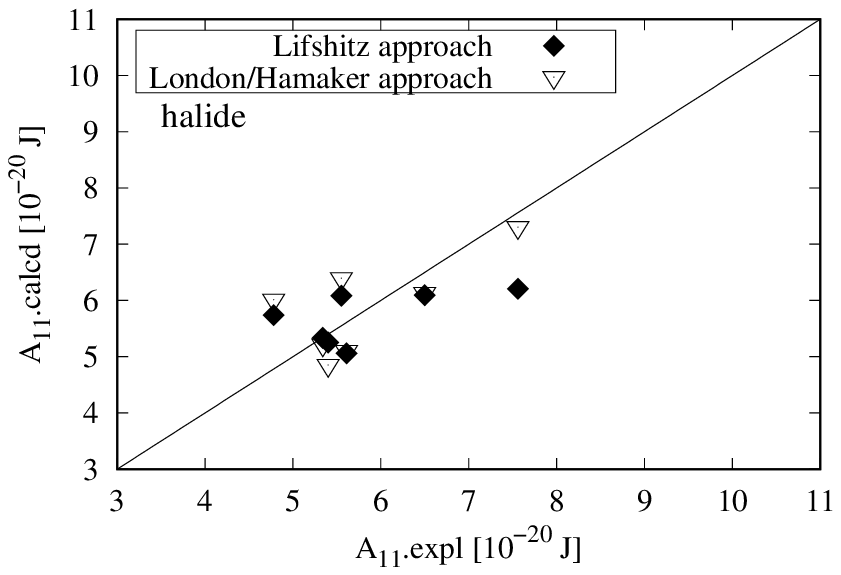}
\vspace{10mm}
\caption{Plots of the calculated values of the Hamaker constant ($A_{11.calcd}$) versus the corresponding experimental values ($A_{11.expl}$) for non-associative materials (59 compounds, top), associative material (11 compounds, middle), and halides (7 compounds, bottom). Please note that the $A_{11.expl}$ values were taken from literature based on the Lifshitz theory (see Supporting information for details).}
\label{fig.HamakerConst3}
\end{figure}

The values for the optical parameters $C_{UV}$, $\omega_{UV}$, $C_{IR}$, and $\omega_{IR}$, and for the Hamaker constant $A_{11}$ were calculated via Eq(\ref{eq.Lifshitz_appr})-Eq(\ref{eq.ixi}). 
For comparison, the values of the Hamaker constants were also calculated by the London/Hamaker approach. 
The values for the ionization potential $I$ required in Eq(\ref{eq.microscopic}) were calculated from the difference between the total electronic energy of the cationic and neutral molecule. 
As can be evinced from the data in Figure \ref{fig.ionizationPotential}, $I$ could be estimated correctly using the B3LYP/aug-cc-pVDZ combination. 
In Figure \ref{fig.HamakerConst3} are reported data that reflect the correlation between the calculated values of the Hamaker constant ($A_{11.calcd}$) and the corresponding experimentally determined ones ($A_{11.expl}$). 
When the London/Hamaker approach was employed, only the $A_{11.calcd}$ values of non-polar small hydrocarbons reproduced the corresponding $A_{11.expl}$ values, whereas in the case of polar materials (e.g., ethers and carbonyls) or large molecules (e.g., hexadecane and bicyclohexyl) $A_{11.expl}$ values were significantly lower than the corresponding $A_{11.expl}$ values. 
As is well known, the London formula is applicable only to non-polar, spherical molecules characterized by diameters smaller than 0.5 nm.\cite{2011ISR} 
The large difference between the $A_{11.expl}$ values and the $A_{11.calcd}$ values determined via the London/Hamaker approach indicates that it is inappropriate to apply the London formula to polar or large molecules. 
By contrast, when applying the Lifshitz approach in this study, the $A_{11.calcd}$ values determined for non-associative materials and halides were close to their $A_{11.expl}$ counterparts. 
Furthermore, through this approach the calculation error for most materials was under 10\%. 
Fortunately, therefore, although the calculation accuracy of the values of the relative dielectric constant was low, the negative consequences on the accuracy of the calculation of the Hamaker constants were small. 
Notably, in fact, in the case of non-associative materials or halides, the contribution of $\epsilon_r$ to $A_{11}$ is relatively small.

With respect to associative materials like alcohols, however, a tendency to underestimating the values of the Hamaker constants $A_{11}$ by this Lifshitz approach was observed. 
This observation was particularly relevant in the case of molecules for which the value of the dielectric constant was significantly underestimated by the calculation, like ethylene glycol or diethylene glycol. 
In these cases, the calculated values of the optical parameter $C_{IR}$ were found to be much smaller than their experimentally determined counterparts, resulting in an underestimation of the values of the Hamaker constant $A_{11}$. 
As mentioned above, use of Kirkwood's equation is required to overcome this problem. 
However, since this equation cannot be solved using a single molecular model, with more complicated procedures required, we did not make use of this equation in the present study.

\begin{table} [htbp]
\caption{Mean absolute relative errors in the calculated values of specific parameters for a range of organic materials. Please note that $n_{589}$ is the refractive index at $\lambda$ = 598 nm. Please note also that the mean absolute relative error in the calculated value of $A_{11}$ was not calculated for carboxylic acid and amines, because no relevant experimental values exist.}
\label{tab.MARE_all}
\begin{center}
\begin{tabular} {  c c c c c c c } \hline 
 & $\alpha$ & $\mu$ & $d$ & $n_{589}$ & $\epsilon_r$ & $A_{11}$ \\ \hline
hydrocarbon & 0.03  & 0.28  & 0.04  & 0.02  & 0.03  & 0.06  \\ 
ether & 0.06  & 0.04  & 0.04  & 0.02  & 0.08  & 0.06  \\ 
ketone, aldehyde & 0.03  & 0.09  & 0.03  & 0.02  & 0.29  & 0.06  \\ 
carboxylic acid & 0.01  & 0.08  & 0.03  & 0.02  & 0.74  &  n/a \\ 
ester & 0.05  & 0.11  & 0.04  & 0.03  & 0.40  & 0.05  \\ 
nitrile & 0.07  & 0.06  & 0.03  & 0.01  & 0.50  & 0.04  \\ 
hydrosilane & 0.13  & 0.02  & 0.04  & 0.05  & 0.10  & 0.11  \\ 
alcohol & 0.04  & 0.08  & 0.04  & 0.02  & 0.62  & 0.24  \\ 
amine & 0.04  & 0.08  & 0.04  & 0.03  & 0.19  &  n/a \\ 
amide & 0.02  & 0.07  & 0.02  & 0.02  & 0.38  & 0.13  \\ 
halide & 0.09  & 0.09  & 0.05  & 0.02  & 0.33  & 0.10  \\ \hline
average & 0.05  & 0.09  & 0.04  & 0.03  & 0.33  & 0.09  \\ 
\hline
\end{tabular}
\end{center}
\end{table}

\section{Conclusion}
In this study, the Hamaker constants $A_{11}$ of organic materials were calculated based on the Lifshitz macroscopic approach performed with a simple DFT calculation. 
The values for the physical properties (polarizability, dipole moment, molecular volume, and vibrational frequency) of organic molecules were computationally estimated by an approach based on the use of the B3LYP density function and of the aug-cc-pVDZ basis set; the relevant Hamaker constants $A_{11}$ were also computationally estimated, in this case using the approximation of the Lorentz-Lorenz equation and Onsager's equation with these properties. 
Employing this approach, we were able to computationally obtain values for the Hamaker constants that were similar to their experimentally determined counterparts for non-associative materials and halides. 
Since this approach does not require the definition of complicated molecular models, even beginners can easily implement it, and the CPU time necessary for it will not be long in comparison with that needed in the conventional London/Hamaker approach. 
The method we developed is especially useful for calculating the Hamaker constants of toxic, degradable, expensive, or rare materials. 
This method is also useful for estimating the Hamaker constants of very large numbers of materials for screening purposes. 
Notably, however, the accuracy of the calculated parameter values is low when this method is applied to associative materials like alcohols; the solution of this problem is a future goal of the research in this field.

\section{acknowledgement}
The computation in this work was performed using the facilities at the Research Center for Advanced Computing Infrastructure (RCACI) at JAIST.
R.M. is grateful for financial support from MEXT-KAKENHI (17H05478 and 16KK0097), from Toyota Motor Corporation, from I-O DATA Foundation, and from the Air Force Office of Scientific Research (AFOSR-AOARD/FA2386-17-1-4049).
K.H. is grateful for financial support from a KAKENHI grant (JP17K17762), a Grant-in-Aid for Scientific Research on Innovative Areas ``Mixed Anion'' Project (JP16H06439) from MEXT, PRESTO (JPMJPR16NA), and the Materials Research by Information Integration Initiative (MI$^2$I) Project of the Support Program for Starting Up Innovation Hub from the Japan Science and Technology Agency (JST). 
R.M. and K.H. are also grateful for financial support from MEXT-FLAGSHIP2020 (hp170269, hp170220).
T. M is grateful for financial support from KAKENHI grant (17H04923 and 17K06013).
The authors would like to thank Enago (www.enago.jp) for the English language review.

\section{supporting information}
The following files are available free of charge.
\begin{itemize}
  \item 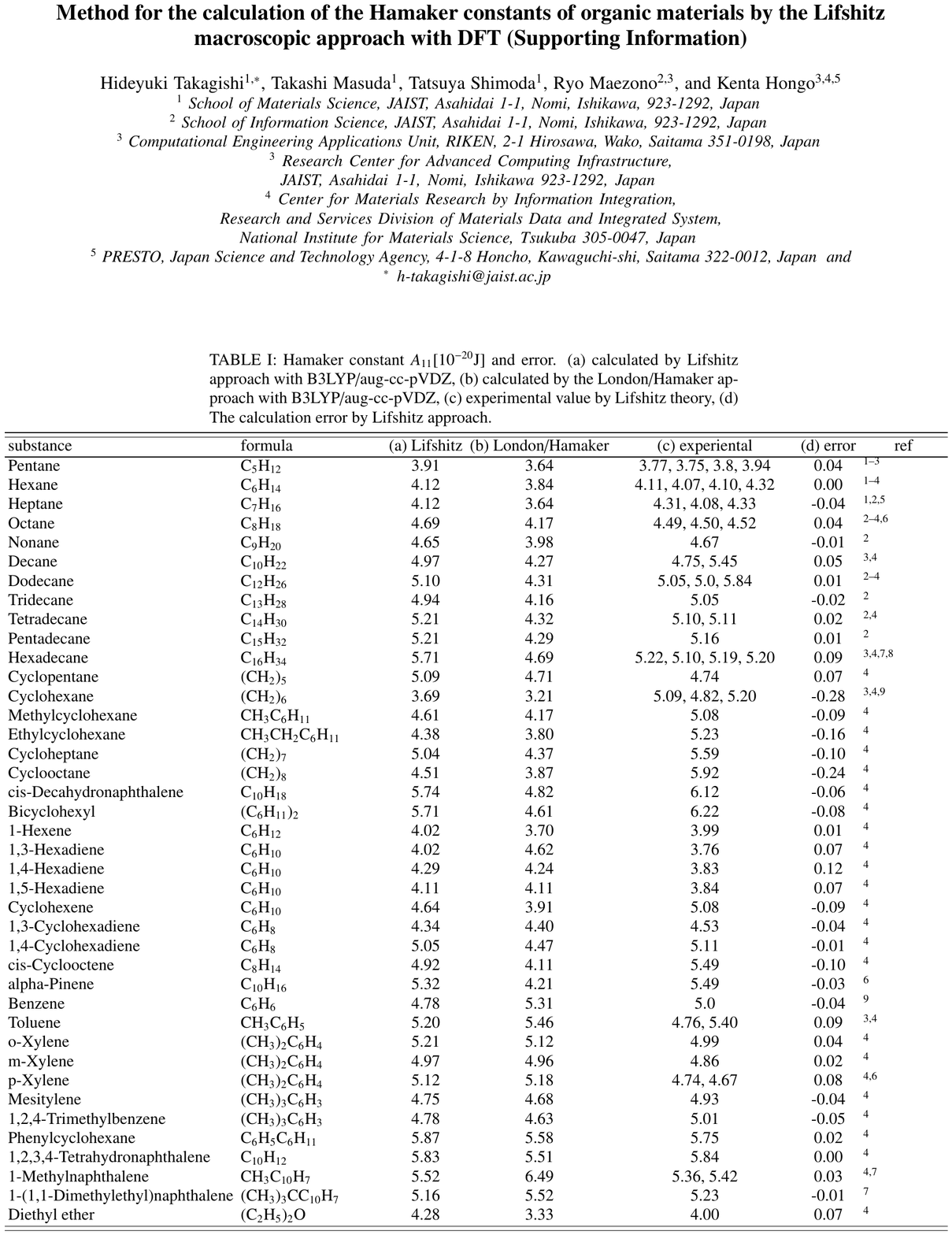: Calculated and experimental values of Hamaker constant for organic materials.
\end{itemize}


\bibliography{references}

\end{document}